# Real-time digital signal recovery for a low-pass transfer function system with multiple complex poles

Jhinhwan Lee

*Department of Physics, Korea Advanced Institute of Science and Technology, Daejeon 34141, Korea*

In order to solve the problems of waveform distortion and signal delay by many physical and electrical systems with linear low-pass transfer characteristics with multiple complex poles, a general digital-signal-processing (DSP)-based method of real-time recovery of the original source waveform from the distorted output waveform is proposed. From the convolution kernel representation of a multiple-pole low-pass transfer function with an arbitrary denominator polynomial with real valued coefficients, it is shown that the source waveform can be accurately recovered in real time using a particular moving average algorithm with real-valued DSP computations only, even though some or all of the poles are complex. The proposed digital signal recovery method is DC-accurate and unaffected by initial conditions, transient signals, and resonant amplitude enhancement. The noise characteristics of the data recovery shows inverse of the low-pass filter characteristics. This method can be applied to most sensors and amplifiers operating close to their frequency response limits or around their resonance frequencies to accurately deconvolute the multiple-pole characteristics and to improve the overall performances of data acquisition systems and digital feedback control systems.

**Introduction**

Many sensors and amplifiers suffer from delayed and distorted responses with single- or multi-pole low-pass characteristics when operated close to their frequency response limits or resonance frequencies. Notable examples of multi-pole systems include a spring-mass system driven by external force or displacement (Fig. 1(a)), or an L-R-C-based circuit driven by external voltage or current (Fig. 1(b)). Unlike a first-order-response system, a second or higher order system may have complex-conjugate-paired poles even though all the coefficients of the denominator polynomial are real. This condition may induce resonant underdamped transfer characteristics and requires a carefully thought-out digital signal recovery algorithm in a conventional DSP system. Here I propose a real-time numerical waveform recovery method that is suitable for the case of arbitrary order denominator polynomials with multiple complex poles and discuss their detailed implementation method, accuracy and noise characteristics.

**Generalization to high-order real systems with complex poles**

Let's consider the two-pole low-pass system as shown in Fig. 1(a) driven by an external displacement to one end of the spring while the output of this system is the displacement of the mass. In case of small enough damping parameter $b$, the two poles can be complex and a complex-conjugate pair. This condition makes the previous method with real-valued poles [1] becomes inappropriate due to the requirement of complex-valued calculations (i.e. $\gamma_i = e^{-s_i T}$ are complex for complex $s_i$) in the real-time core of the DSP/FPGA signal processor.

Here I will introduce an appropriate digital signal recovery method for a general two-pole linear low-pass system by direct mathematical expansion starting from the convolution kernel expression of the two pole transfer characteristics and show that it is equivalent to the nested multi-pole solution shown in Fig. 4 of Ref. [1]. The results will then be generalized into higher order low-pass system whose Laplace transfer function has an $n$-th order real-coefficient polynomial as its denominator.

The Laplace representation of the first order low pass characteristics *with an initial condition* (Ref. [2])

$$V_o(s) = \frac{1}{1+s/s_1} V_i(s) + \frac{1}{s+s_1} V_i(t=0) \tag{1}$$

with time constant $\tau_1 = 1/s_1$ can be converted to the time domain expression of

$$V_o(t) = \int_0^\infty dt'\, s_1 e^{-s_1 t'} V_i(t-t') + V_i(t=0) e^{-s_1 t} \Theta(t). \tag{2}$$

We can recover the original signal as derived in Ref. [1]

$$V_i\left(t-\frac{T}{2}\right) \approx \frac{V_o(t) - e^{-s_1 T} V_o(t-T)}{1 - e^{-s_1 T}} \tag{3}$$

since the contribution from the initial condition term $V_o^t(t) = V_i(t=0)e^{-s_1 t}\Theta(t)$ vanishes as

$$\frac{V_o^t(t) - e^{-s_1 T}V_o^t(t-T)}{1 - e^{-s_1 T}} = V_i(t=0)\frac{e^{-s_1 t}\Theta(t) - e^{-s_1 T}e^{-s_1(t-T)}\Theta(t-T)}{1 - e^{-s_1 T}} = 0 \tag{4}$$

for arbitrary $t > T$.

Now let's consider the general second order low-pass characteristics with initial conditions

$$V_o(s) = \frac{1}{1+bs+as^2}V_i(s) + \frac{b+as}{1+bs+as^2}V_i(t=0) + \frac{a}{1+bs+as^2}V'_i(t=0) \tag{5}$$

$$= \frac{1}{\left(1+\frac{s}{s_1}\right)\left(1+\frac{s}{s_2}\right)}V_i(s) + \frac{1}{s+s_1}d_1 + \frac{1}{s+s_2}d_2. \tag{6}$$

This can be converted to the time domain solution of

$$V_o(t) = V_o^0(t) + V_o^t(t) \tag{7}$$

which is the sum of the nested time domain convolution

$$V_o^0(t) = \int_0^\infty dt''\, s_2 e^{-s_2 t''} \int_0^\infty dt'\, s_1 e^{-s_1 t'} V_i(t-t'-t'') \tag{8}$$

and the transient solution dependent on the initial conditions

$$V_o^t(t) = d_1 e^{-s_1 t}\Theta(t) + d_2 e^{-s_2 t}\Theta(t). \tag{9}$$

Since

$$V_o^0(t-T) = \int_0^\infty dt''\, s_2 e^{-s_2 t''} \int_0^\infty dt'\, s_1 e^{-s_1 t'} V_i(t-T-t'-t'') \tag{10}$$

$$= e^{s_2 T} \int_T^\infty dt''\, s_2 e^{-s_2 t''} \int_0^\infty dt'\, s_1 e^{-s_1 t'} V_i(t-t'-t'') \tag{11}$$

$$= e^{s_1 T} \int_0^\infty dt''\, s_2 e^{-s_2 t''} \int_T^\infty dt'\, s_1 e^{-s_1 t'} V_i(t-t'-t'') \tag{12}$$

and

$$V_o^0(t-2T) = e^{s_1 T}e^{s_2 T} \int_T^\infty dt''\, s_2 e^{-s_2 t''} \int_T^\infty dt'\, s_1 e^{-s_1 t'} V_i(t-t'-t''), \tag{13}$$

we have

$$\int_0^T dt''\, s_2 e^{-s_2 t''} \int_0^T dt'\, s_1 e^{-s_1 t'} V_i(t-t'-t'') = V_o^0(t) - e^{-s_1 T}V_o^0(t-T) - e^{-s_2 T}V_o^0(t-T) + e^{-s_1 T}e^{-s_2 T}V_o^0(t-2T) \tag{14}$$

which is graphically illustrated in Fig. 2(d)-(h).

In the limit of small $T$, the average value of $V_i$ over the square region of integration with parameter range $[t-2T, t]$ can be approximated by $V_i(t-T)$ and the left-hand side expression is simplified as

$$\int_0^T dt''\, s_2 e^{-s_2 t''} \int_0^T dt'\, s_1 e^{-s_1 t'} V_i(t-t'-t'') \approx V_i(t-T) \int_0^T dt''\, s_2 e^{-s_2 t''} \int_0^T dt'\, s_1 e^{-s_1 t'} \tag{15}$$

$$= V_i(t-T)(1 - e^{-s_1 T})(1 - e^{-s_2 T}) \tag{16}$$

Therefore we have

$$V_i(t-T) \approx \frac{V_o^0(t) - \{e^{-s_1 T} + e^{-s_2 T}\}V_o^0(t-T) + e^{-s_1 T}e^{-s_2 T}V_o^0(t-2T)}{(1 - e^{-s_1 T})(1 - e^{-s_2 T})}. \tag{17}$$

Since for both $i = 1, 2$ we have

$$e^{-s_i t} - \{e^{-s_1 T} + e^{-s_2 T}\}e^{-s_i(t-T)} + e^{-s_1 T}e^{-s_2 T}e^{-s_i(t-2T)} = 0, \tag{18}$$

the contribution from the initial-condition-dependent transient solution $V_o^t(t)$ vanishes for arbitrary $t > 2T$ when inserted in the places of the $V_o^0(t)$:

$$\frac{V_o^t(t) - \{e^{-s_1 T} + e^{-s_2 T}\}V_o^t(t-T) + e^{-s_1 T}e^{-s_2 T}V_o^t(t-2T)}{(1 - e^{-s_1 T})(1 - e^{-s_2 T})} = 0. \tag{19}$$

Therefore the signal recovery formula for $V_o(t)$ maintains the same form as the Eq. (17):

$$V_i(t-T) \approx \frac{V_o(t) - \{e^{-s_1 T} + e^{-s_2 T}\}V_o(t-T) + e^{-s_1 T}e^{-s_2 T}V_o(t-2T)}{(1 - e^{-s_1 T})(1 - e^{-s_2 T})}. \tag{20}$$

We can also show that the above Eq. (20) can be converted to a nested form

$$V_i(t-T) \approx \frac{\left\{\frac{V_o(t) - e^{-s_1 T}V_o(t-T)}{1 - e^{-s_1 T}}\right\} - e^{-s_2 T}\left\{\frac{V_o(t-T) - e^{-s_1 T}V_o(t-2T)}{1 - e^{-s_1 T}}\right\}}{1 - e^{-s_2 T}} \tag{21}$$

$$= \frac{V_{oi}(t) - e^{-s_2 T}V_{oi}(t-T)}{1 - e^{-s_2 T}} \tag{22}$$

where we define the intermediately recovered waveform

$$V_{oi}(t) = \frac{V_o(t) - e^{-s_1 T}V_o(t-T)}{1 - e^{-s_1 T}}. \tag{23}$$

This shows clearly that the single register implementation of the Eq. (20) is equivalent to the cascaded two register implementation shown in Fig. 4 of Ref. [1].

The above result can be generalized to an arbitrary high order $n$. The general $n$-th order low-pass characteristics with initial conditions is given by

$$V_o(s) = \frac{1}{\sum_{i=0}^{n} a_i s^i} V_i(s) + \frac{\sum_{i=1}^{n} \sum_{j=1}^{i} a_i s^{i-j} V_i^{(j-1)}(t=0)}{\sum_{i=0}^{n} a_i s^i} \tag{24}$$

$$= \frac{1}{\prod_{i=1}^{n}\left(1+\frac{s}{s_i}\right)} V_i(s) + \sum_{i=1}^{n} \frac{1}{s+s_i} d_i \tag{25}$$

where $d_i$ is a linear combination of the initial conditions $V_i^{(0,1,2,\ldots,n-1)}(t=0)$ [2].

The time-domain solution is given in the form

$$V_o(t) = V_o^0(t) + V_o^t(t) \tag{26}$$

where

$$V_o^0(t) = \int_0^\infty dt_n\, s_n e^{-s_n t_n} \cdots \int_0^\infty dt_2\, s_2 e^{-s_2 t_2} \int_0^\infty dt_1\, s_1 e^{-s_1 t_1} V_i(t - t_1 - t_2 - \cdots - t_n) \tag{27}$$

and

$$V_o^t(t) = \sum_{i=1}^{n} d_i e^{-s_i t} \Theta(t). \tag{28}$$

The signal recovery formula (corresponding to Eqs. (12) and (20)) is generalized to

$$V_i\left(t - \frac{T}{2} n\right) \approx$$
$$\frac{V_o(t) - \{e^{-s_1 T} + e^{-s_2 T} + \cdots + e^{-s_n T}\} V_o(t-T) + \{e^{-s_1 T} e^{-s_2 T} + e^{-s_1 T} e^{-s_3 T} + \cdots + e^{-s_{n-1} T} e^{-s_n T}\} V_o(t-2T) + \cdots + (-1)^n \{e^{-s_1 T} e^{-s_2 T} e^{-s_3 T} \cdots e^{-s_n T}\} V_o(t-nT)}{(1-e^{-s_1 T})(1-e^{-s_2 T})(1-e^{-s_3 T}) \cdots (1-e^{-s_n T})} \tag{29}$$

where the coefficients for $V_o(t - mT)$ in the numerator are simply the coefficients for the $z^{n-m}$ term in the generating polynomial

$$F(z) = (z - e^{-s_1 T})(z - e^{-s_2 T})(z - e^{-s_3 T}) \cdots (z - e^{-s_n T}). \tag{30}$$

We can again show that the transient solution $V_o^t(t)$ gives no contribution to the right-hand side of the Eq. (29) since, for the arbitrary $i$-th term of the Eq. (28) proportional to $e^{-s_i t}$, we have

$$e^{-s_i t} - \{e^{-s_1 T} + e^{-s_2 T} + \cdots + e^{-s_n T}\} e^{-s_i(t-T)} + \{e^{-s_1 T} e^{-s_2 T} + e^{-s_1 T} e^{-s_3 T} + \cdots + e^{-s_{n-1} T} e^{-s_n T}\} e^{-s_i(t-2T)}$$
$$+ \cdots + (-1)^n \{e^{-s_1 T} e^{-s_2 T} e^{-s_3 T} \cdots e^{-s_n T}\} e^{-s_i(t-nT)}$$
$$= e^{-s_i(t-nT)} \begin{bmatrix} e^{-n s_i T} - \{e^{-s_1 T} + e^{-s_2 T} + \cdots + e^{-s_n T}\} e^{-(n-1)s_i T} \\ + \{e^{-s_1 T} e^{-s_2 T} + e^{-s_1 T} e^{-s_3 T} + \cdots + e^{-s_{n-1} T} e^{-s_n T}\} e^{-(n-2)s_i T} + \cdots + (-1)^n \{e^{-s_1 T} e^{-s_2 T} e^{-s_3 T} \cdots e^{-s_n T}\} \end{bmatrix} \tag{31}$$

$$= e^{-s_i(t-nT)} F(e^{-s_i T}) = 0 \tag{32}$$

and the right-hand side of the Eq. (29) vanishes when $V_o(t)$ is replaced by $V_o^t(t)$.

This provides a general proof that the signal recovery method shown in the Eq. (29) produces output signal completely independent of the initial conditions and the transient signals.

**Device Implementation**

Fig. 3 shows three equivalent representations for physical implementation of high order signal recovery in the DSP/FPGA device. They are mathematically equivalent but in case of complex poles, the first implementation (Fig. 3(a)) requires complex computation while the second (Fig. 3(b)) and the third (Fig. 3(c)) require only real-valued computation which has a significant advantage in the real-time process core of the DSP/FPGA.

We note that the complex roots of any real-coefficient polynomial always occur in complex-conjugated pairs. Therefore in the device implementation of Fig. 3(b) (formed by combining every pair of register loops of Fig. 3a for a complex conjugate pair (such as $s_1$ and $s_2$ with $D = b^2 - 4a < 0$) into a single register loop), it is sufficient to show that only real-valued computations are needed for the evaluation of the combined second order Eq. (20).

Let's assume that $s_1 = \alpha + i\beta$ and $s_2 = \alpha - i\beta$ where $\alpha$ and $\beta$ are real and $\alpha > 0$. Then we have the following values all real:

$$e^{-s_1 T} + e^{-s_2 T} = e^{-(\alpha+i\beta)T} + e^{-(\alpha-i\beta)T} = 2 e^{-\alpha T} \cos \beta T \tag{33}$$

$$e^{-s_1 T} e^{-s_2 T} = e^{-(\alpha+i\beta)T - (\alpha-i\beta)T} = e^{-2\alpha T} \tag{34}$$

and all the numbers used in the evaluation of the Eq. (20) become real

$$V_i(t - T) \approx \frac{V_o(t) - \{2 e^{-\alpha T} \cos \beta T\} V_o(t-T) + e^{-2\alpha T} V_o(t-2T)}{1 - 2 e^{-\alpha T} \cos \beta T + e^{-2\alpha T}}. \tag{35}$$

The device implementation of Fig. 3(c) (formed by combining all the register loops of Fig. 3a into one) also has the property of real-value-only computations. The proof that all the coefficients in the numerator of the Eq. (29) and its denominator are real-valued is easily provided noting that if $s_1$ and $s_2$ are a complex-conjugate pair, $e^{-s_1 T}$ and $e^{-s_2 T}$ are also. As a result, all the coefficients of the polynomial expansion of the Eq. (30) are real. Therefore, all the coefficients appearing in the numerator of Eq. (29) are real and the denominator $F(1)$ is real.

**Noise consideration**

In order to understand the noise characteristics of the multi-pole recovery method quantitatively, let's assume without too much loss of generality, that the high-order-convoluted analog output signal $\tilde{V}_O(t)$ contains a slow-varying (compared to $2T$) raw signal $V_O(t)$ plus a pseudo-random noise $V_n(t)$ whose correlation time is shorter than $T$. First, let's look at the nontrivial second-order case of Eq. (35) where the two roots $s_1$ and $s_2$ are a complex-conjugated pair. Then the numerical recovery operation applied to $\tilde{V}_O(t) = V_O(t) + V_n(t)$ can be divided into two terms

$$V_R(t-T) \approx \frac{V_O(t) - \{2e^{-\alpha T}\cos\beta T\}V_O(t-T) + e^{-2\alpha T}V_O(t-2T)}{1 - 2e^{-\alpha T}\cos\beta T + e^{-2\alpha T}} + \frac{V_n(t) - \{2e^{-\alpha T}\cos\beta T\}V_n(t-T) + e^{-2\alpha T}V_n(t-2T)}{1 - 2e^{-\alpha T}\cos\beta T + e^{-2\alpha T}} \quad (36)$$

where the first term gives the slow varying signal with value approximated by $V_O(t)(\approx V_O(t-T) \approx V_O(t-2T))$ and the second term gives noise level proportional to $\frac{\sqrt{1+(2e^{-\alpha T}\cos\beta T)^2+e^{-4\alpha T}}}{1-2e^{-\alpha T}\cos\beta T+e^{-2\alpha T}}|V_n(t)|$ due to the presumed absence of time correlation between noise $V_n(t)$, $V_n(t-T)$, and $V_n(t-2T)$. For small $\alpha T \ll 1$ and $\beta T \ll 1$, the signal-to-noise (S/N) ratio is reduced by a factor of

$$\frac{SN}{SN_0} \approx \frac{1-2e^{-\alpha T}\cos\beta T+e^{-2\alpha T}}{\sqrt{1+(2e^{-\alpha T}\cos\beta T)^2+e^{-4\alpha T}}} \approx \frac{(\alpha^2+\beta^2)T^2}{\sqrt{6}} = \frac{s_1 s_2 T^2}{\sqrt{6}} \ll 1. \quad (37)$$

This is formally identical to the result when the two poles are real

$$\frac{SN}{SN_0} \approx \frac{(1-e^{-s_1 T})(1-e^{-s_2 T})}{\sqrt{1+(e^{-s_1 T}+e^{-s_2 T})^2+(e^{-s_1 T}e^{-s_2 T})^2}} \approx \frac{s_1 s_2 T^2}{\sqrt{6}}. \quad (38)$$

Generalization to an arbitrarily high order case of Eq. (29) leads to

$$\frac{SN}{SN_0} \approx \frac{(1-e^{-s_1 T})(1-e^{-s_2 T})(1-e^{-s_3 T})\cdots(1-e^{-s_n T})}{\sqrt{1+\{e^{-s_1 T}+e^{-s_2 T}+\cdots+e^{-s_n T}\}^2+\{e^{-s_1 T}e^{-s_2 T}+e^{-s_1 T}e^{-s_3 T}+\cdots+e^{-s_{n-1} T}e^{-s_n T}\}^2+\cdots+\{e^{-s_1 T}e^{-s_2 T}e^{-s_3 T}\cdots e^{-s_n T}\}^2}} \quad (39)$$

$$\approx \frac{|s_1 s_2 s_3 \cdots s_n|T^n}{\sqrt{\{_nC_0\}^2+\{_nC_1\}^2+\{_nC_2\}^2+\cdots+\{_nC_n\}^2}} \quad (40)$$

$$= \frac{|s_1 s_2 s_3 \cdots s_n|T^n}{\sqrt{_{2n}C_n}}. \quad (41)$$

As mentioned in Ref. [1], if we sample $\tilde{V}_O(t)$ $N_s (\geq 2)$ times over the short time intervals within $[t, t-T)$ and use their averages in place of $\tilde{V}_O(t)$, we may further increase the S/N by up to a factor given by a fraction of $\sqrt{N_s}$. The factor can approach $\sqrt{N_s}$ in case the correlation time of the noise is still shorter than the sampling periods of the $N_s$ data points. On the other hand, when it is possible to perform $N_a$ multiple measurements over a repeated input signal, we can increase the S/N to an arbitrary level by choosing the averaging $N_a$ by

$$N_a \geq \left(\frac{SN_{goal}}{SN_0}\right)^2 \frac{_{2n}C_n}{|s_1 s_2 s_3 \cdots s_n|^2 T^{2n}} \quad (42)$$

**Simulated demonstration**

I performed simulations for a two-pole spring-mass-damper system shown in Fig. 1(a) whose results are shown in Fig. 4-6 and for a three-pole passive electrical circuit system (shown in Fig. 1(b)) whose results are shown in Fig. 7-9. In each case, we first numerically solved the (second or third order) differential equations for three different kinds of input waveforms and processed it with the real-time data recovery scheme shown in Fig. 3(c) with optimal and non-optimal choices of parameters. The signal recovery scheme of Fig. 3(b) produced numerically identical results as the scheme of Fig. 3(c) and therefore not separately reproduced here.

As can be seen in all Figs. 4-9, the recovered waveform closely matches with the input waveform only when the parameter $T$ used in evaluating all the coefficients of the Eq. (20) or (29) matches with the actual time difference between samplings, leading to the optimal values of all the coefficients and hence the optimal overall compensation. Also it should be noted that the signal recovery is DC-accurate and independent of the initial conditions, the transient waveforms and the waveforms significantly amplified near the resonance frequency.

## Conclusion

A relatively simple digital-signal-processing-based method of real-time signal recovery is proposed, which can compensate for the waveform distortion and propagation delay due to single-pole or multiple-complex-pole low-pass transfer characteristics in many physical and electronic systems. In case the transfer function has a real-coefficient polynomial as its denominator, we can use signal processing based on real-valued computations only, even though some of the poles are complex. The overall method is also initial-value-independent and will be especially useful in exactly deconvoluting the multi-pole transfer characteristics, in improving the performances of data acquisition systems and in stabilizing high speed feedback control systems with sensors and amplifiers operated close to their frequency response limits or around their resonance frequencies, utilizing modern low-cost high-speed DSPs and FPGAs [3-9].

## Acknowledgements

This work was supported by the Pioneer Research Center Program (No. NRF-2013M3C1A3064455), the Basic Science Research Program (No. NRF-2017R1D1A1B01016186) and the Brain Korea 21 Plus Program through the NRF of Korea.

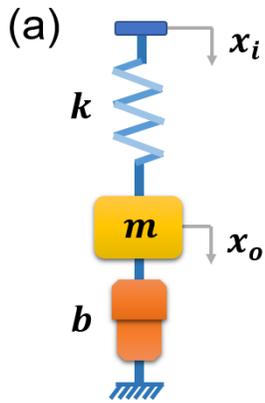
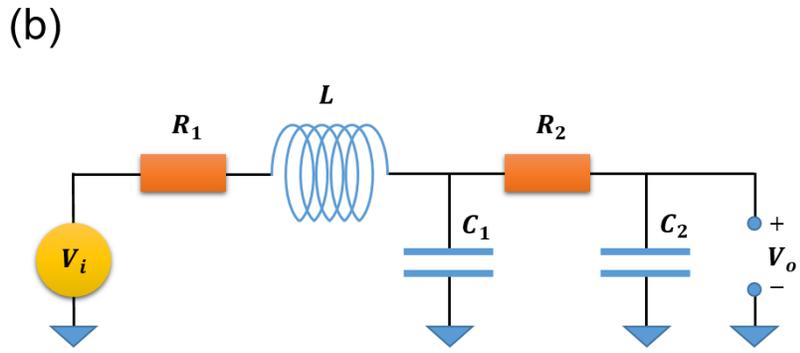

$$\frac{x_o}{x_i} = \frac{1}{1 + \frac{b}{k}s + \frac{m}{k}s^2}$$

$$\frac{V_o}{V_i} = \frac{1}{1 + (C_1R_1 + C_2R_1 + C_2R_2)s + (C_1L + C_2L + C_1C_2R_1R_2)s^2 + C_1C_2LR_2s^3}$$

Fig. 1. Examples of systems with multiple-pole transfer characteristics also used for simulations in Figs. 4-9. (a) A two-pole low-pass mechanical system made of a series connection of a spring ($k$), a mass ($m$) and a damper ($b$) whose transfer characteristics is used for Figs. 4-6. (b) A three-pole low-pass circuit including two capacitors and one inductor whose transfer characteristics is used for Figs. 7-9.

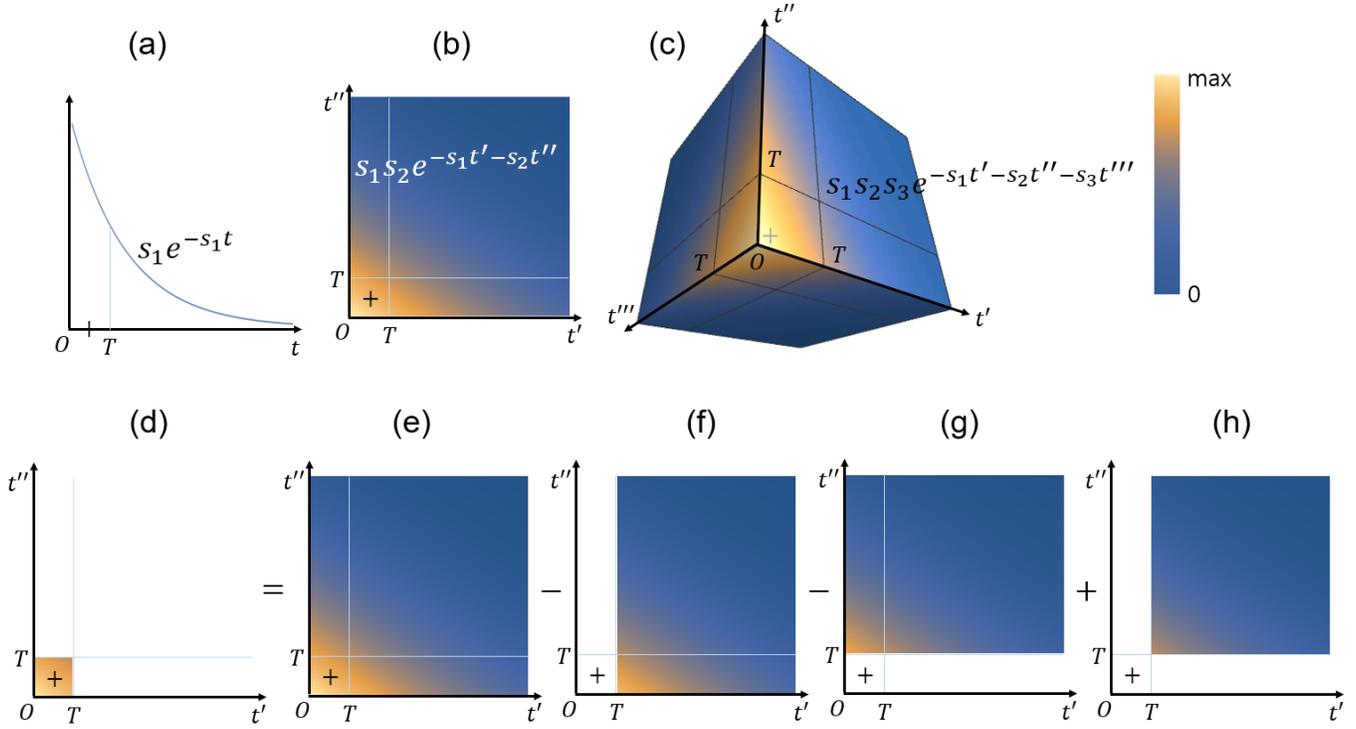

Fig. 2. (a)-(c) Convolution kernel functions for single-pole ($d = 1$) (a), two-pole ($d = 2$) (b) and three-pole ($d = 3$) (c) low-pass systems. We assumed real values for $s_1, s_2, s_3, \ldots$ for simplicity but some of them can be complex conjugated pairs while unpaired ones should be real. The input function's value at the center of the cubical volume $T^d$ can be well approximated by the convolution integral within the cubical volume $T^d$ (that can be evaluated by subtracting and adding convolution integrals with all possible combinations of time domain offsets as illustrated in (d)-(h) for $d = 2$) divided by the kernel-only integral within the same volume. The convolution integrals over the shaded regions in (d)-(h) are as follows. (d) $\int_0^T dt''\, s_2 e^{-s_2 t''} \int_0^T dt'\, s_1 e^{-s_1 t'} V_i(t - t' - t'') \approx V_i(t - T)(1 - e^{-s_1 T})(1 - e^{-s_2 T})$ (e) $\int_0^\infty dt''\, s_2 e^{-s_2 t''} \int_0^\infty dt'\, s_1 e^{-s_1 t'} V_i(t - t' - t'') = V_o(t)$ (f) $e^{s_1 T} \int_0^\infty dt''\, s_2 e^{-s_2 t''} \int_T^\infty dt'\, s_1 e^{-s_1 t'} V_i(t - t' - t'') = e^{-s_1 T} V_o(t - T)$ (g) $e^{s_2 T} \int_T^\infty dt''\, s_2 e^{-s_2 t''} \int_0^\infty dt'\, s_1 e^{-s_1 t'} V_i(t - t' - t'') = e^{-s_2 T} V_o(t - T)$ (h) $e^{s_1 T} e^{s_2 T} \int_T^\infty dt''\, s_2 e^{-s_2 t''} \int_T^\infty dt'\, s_1 e^{-s_1 t'} V_i(t - t' - t'') = e^{-s_1 T} e^{-s_2 T} V_o(t - 2T)$.

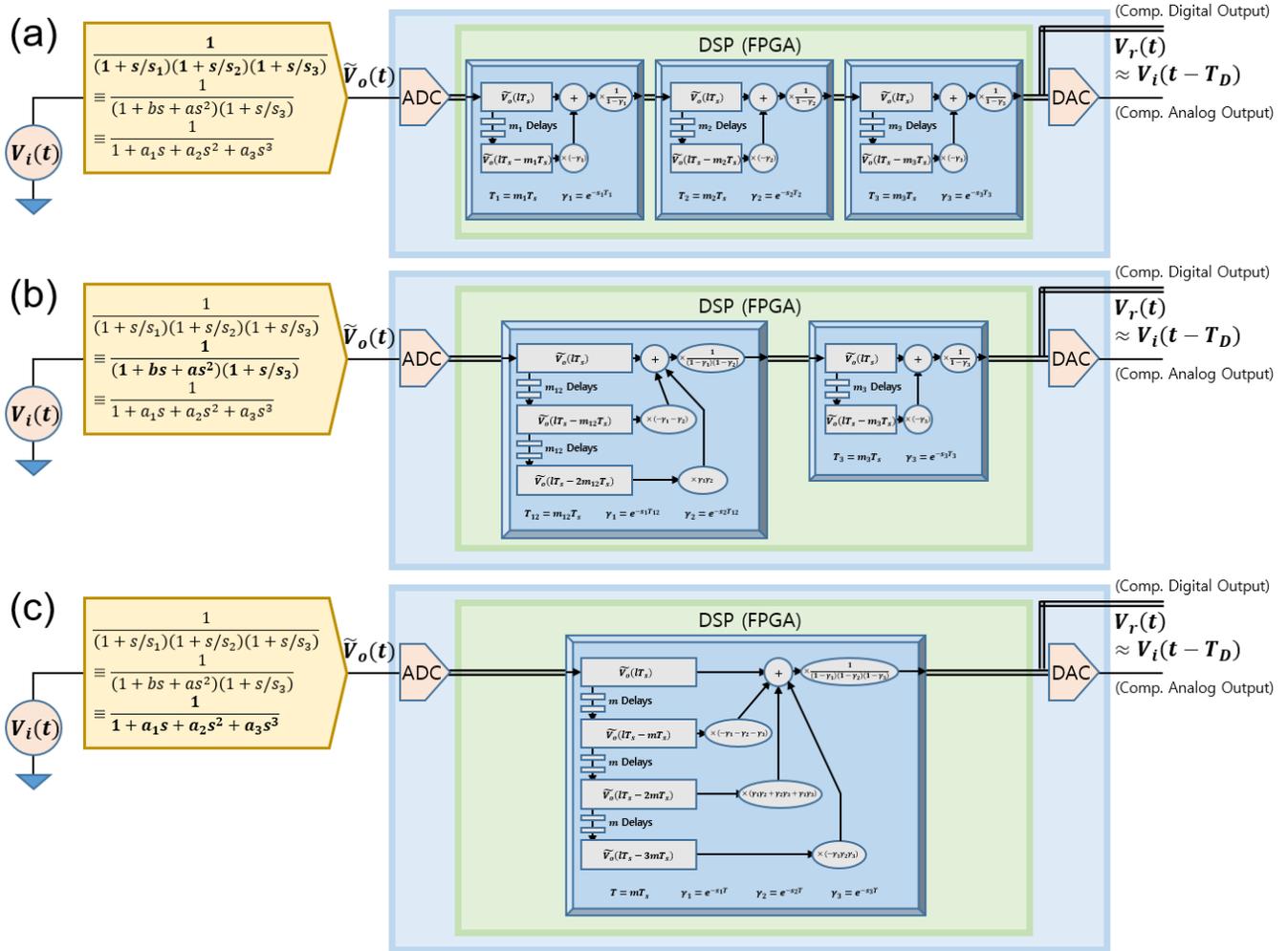

Fig. 3. Schematic diagrams of a real-time digital signal recovery system (in blue boxes on the right) compensating for the signal distortion by a physical or electrical system with multi-pole low-pass transfer characteristics (in yellow pentagons on the left). (a) In case all the $s_i$ are real and positive, the nested multiple register scheme as illustrated in Ref. [1] can be used. However, in case some of the $s_i$ are complex, the scheme is not efficiently realized in real-valued digital signal processors. Two alternative solutions are suggested: (b) Combining two registers for every complex conjugated pair of the $s_i$ (e.g. $D = b^2 - 4a < 0$) into one, while leaving the registers for real $s_i$ left nested as before, all the digital signal processing is done with real parameters and real-valued digital signal processing only. (c) Combining all the registers of (a) into one register also results in the digital signal processing done with real parameters and real-valued digital signal processing only.

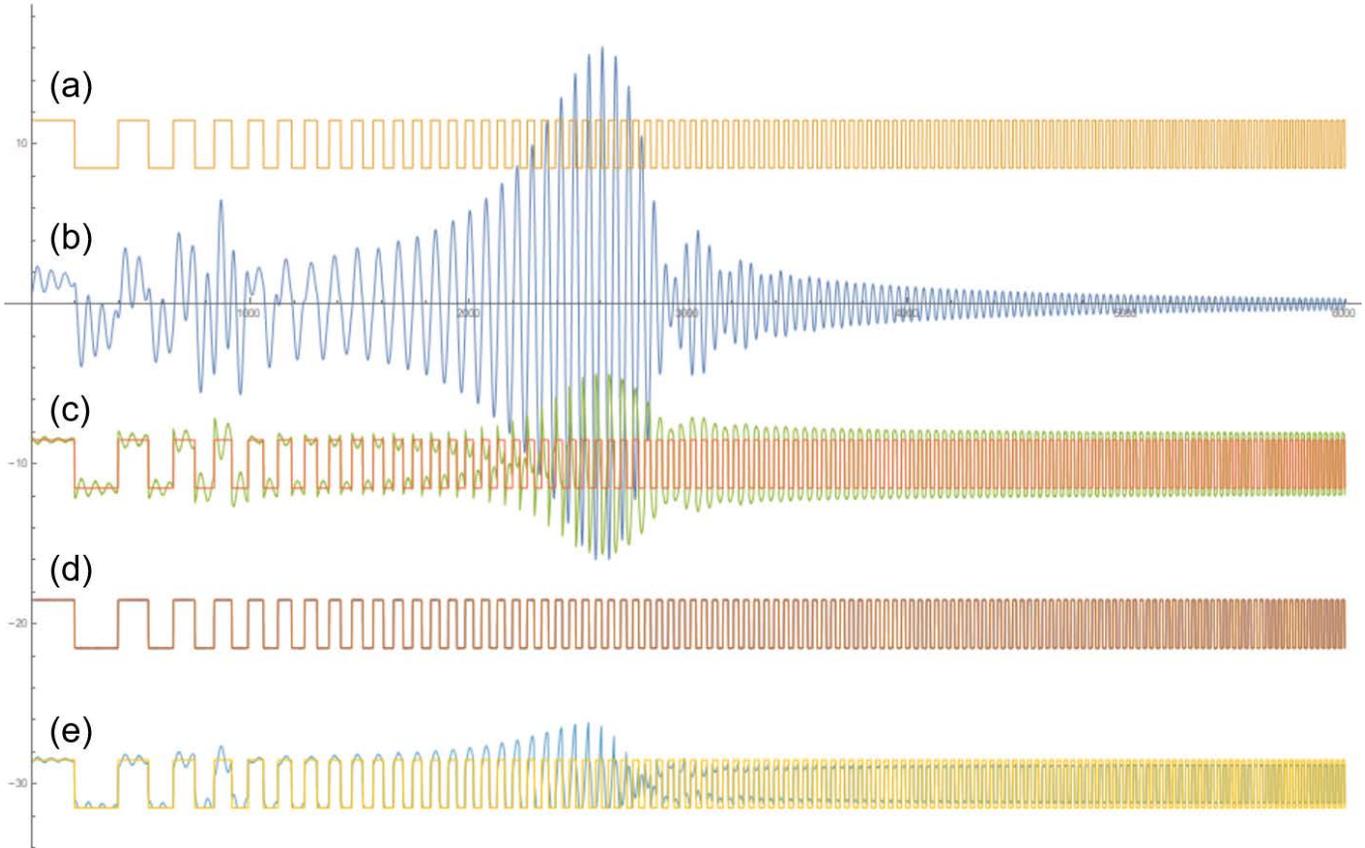

Fig. 4. Numerical simulation of the real-time waveform recovery in the scheme of Fig. 3(c) for a system with two-pole transfer characteristics shown in Fig. 1(a), tested with a frequency-varying rectangular input waveform. (a) Input waveform $x_i(t)$. (b) Response of the system output $x_o(t)$ calculated by numerically solving the differential equation with the two initial conditions $x_o(0)$ and $\dot{x}_o(0)$ arbitrarily chosen. (c) Compensated waveform $x_r(t)$ (overlaid with the input waveform $x_i(t)$) calculated from the waveform in (b) with all the $\gamma_i$ parameters intentionally offset from their optimal values by replacing every $T$ by $0.9T$. (d) Compensated waveform $x_r(t)$ (overlaid with the input waveform $x_i(t)$) calculated from the waveform in (b) with all the $\gamma_i$ parameters set at their optimal values. (e) Compensated waveform $x_r(t)$ (overlaid with the input waveform $x_i(t)$) calculated from the waveform in (b) with all the $\gamma_i$ parameters intentionally offset from their optimal values by replacing every $T$ by $1.1T$. Note that the strong transient waveforms after the sharp transitions and the strong resonant waveform near $2000 < t < 3500$ in (b) are exactly cancelled out in (d).

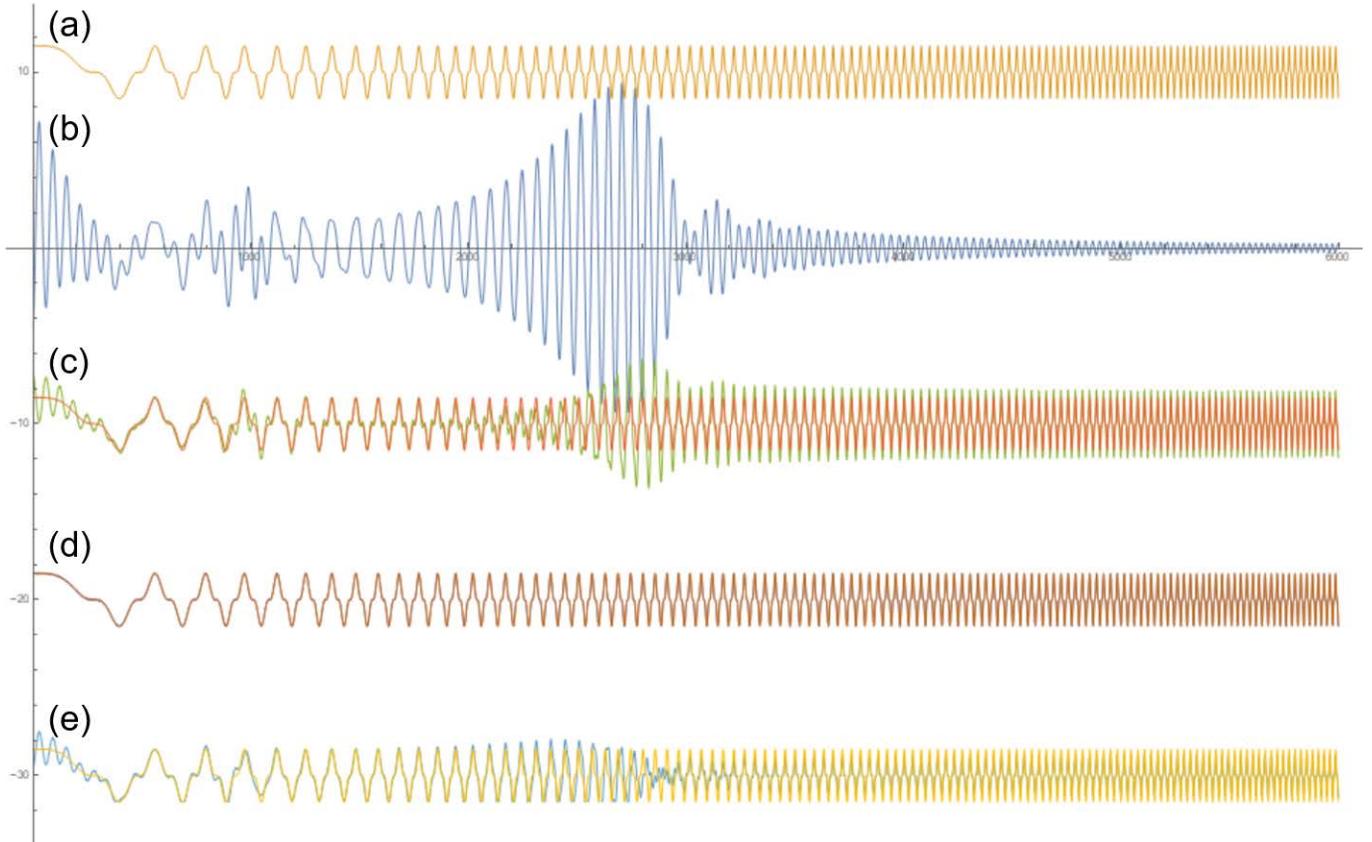

Fig. 5. Numerical simulation of the real-time waveform recovery in the scheme of Fig. 3(c) for a system with two-pole transfer characteristics shown in Fig. 1(a), tested with a frequency-varying cosine-cubed input waveform. (a) Input waveform $x_i(t)$. (b) Response of the system output $x_o(t)$ calculated by numerically solving the differential equation with the two initial conditions $x_o(0)$ and $\dot{x}_o(0)$ arbitrarily chosen. (c) Compensated waveform $x_r(t)$ (overlaid with the input waveform $x_i(t)$) calculated from the waveform in (b) with all the $\gamma_i$ parameters intentionally offset from their optimal values by replacing every $T$ by $0.9T$. (d) Compensated waveform $x_r(t)$ (overlaid with the input waveform $x_i(t)$) calculated from the waveform in (b) with all the $\gamma_i$ parameters set at their optimal values. (e) Compensated waveform $x_r(t)$ (overlaid with the input waveform $x_i(t)$) calculated from the waveform in (b) with all the $\gamma_i$ parameters intentionally offset from their optimal values by replacing every $T$ by $1.1T$. Note that the strong transient waveform near $0 < t < 400$ in (b) and the strong resonant waveform near $2000 < t < 3500$ are exactly cancelled out in (d).

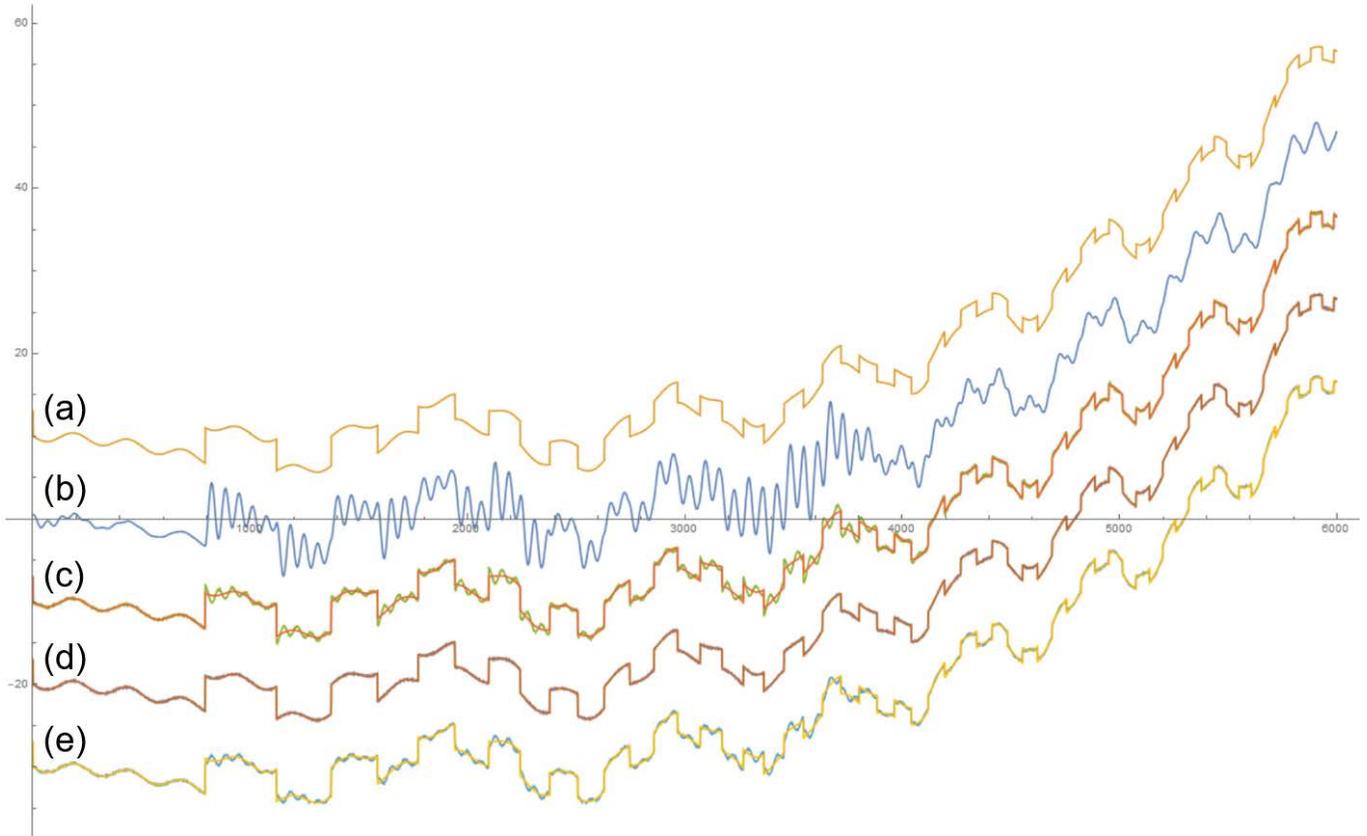

Fig. 6. Numerical simulation of the real-time waveform recovery in the scheme of Fig. 3(c) for a system with two-pole transfer characteristics shown in Fig. 1(a), tested with an aperiodic input waveform. (a) Input waveform $x_i(t)$. (b) Response of the system output $x_o(t)$ calculated by numerically solving the differential equation with the two initial conditions $x_o(0)$ and $\dot{x}_o(0)$ arbitrarily chosen. (c) Compensated waveform $x_r(t)$ (overlaid with the input waveform $x_i(t)$) calculated from the waveform in (b) with all the $\gamma_i$ parameters intentionally offset from their optimal values by replacing every $T$ by $0.9T$. (d) Compensated waveform $x_r(t)$ (overlaid with the input waveform $x_i(t)$) calculated from the waveform in (b) with all the $\gamma_i$ parameters set at their optimal values. (e) Compensated waveform $x_r(t)$ (overlaid with the input waveform $x_i(t)$) calculated from the waveform in (b) with all the $\gamma_i$ parameters intentionally offset from their optimal values by replacing every $T$ by $1.1T$. Note that the strong transient waveforms after the sharp transitions are exactly cancelled out in (d) and the recovery is DC-accurate even with the parameters slightly offset from their optimal values as shown in (c)-(e).

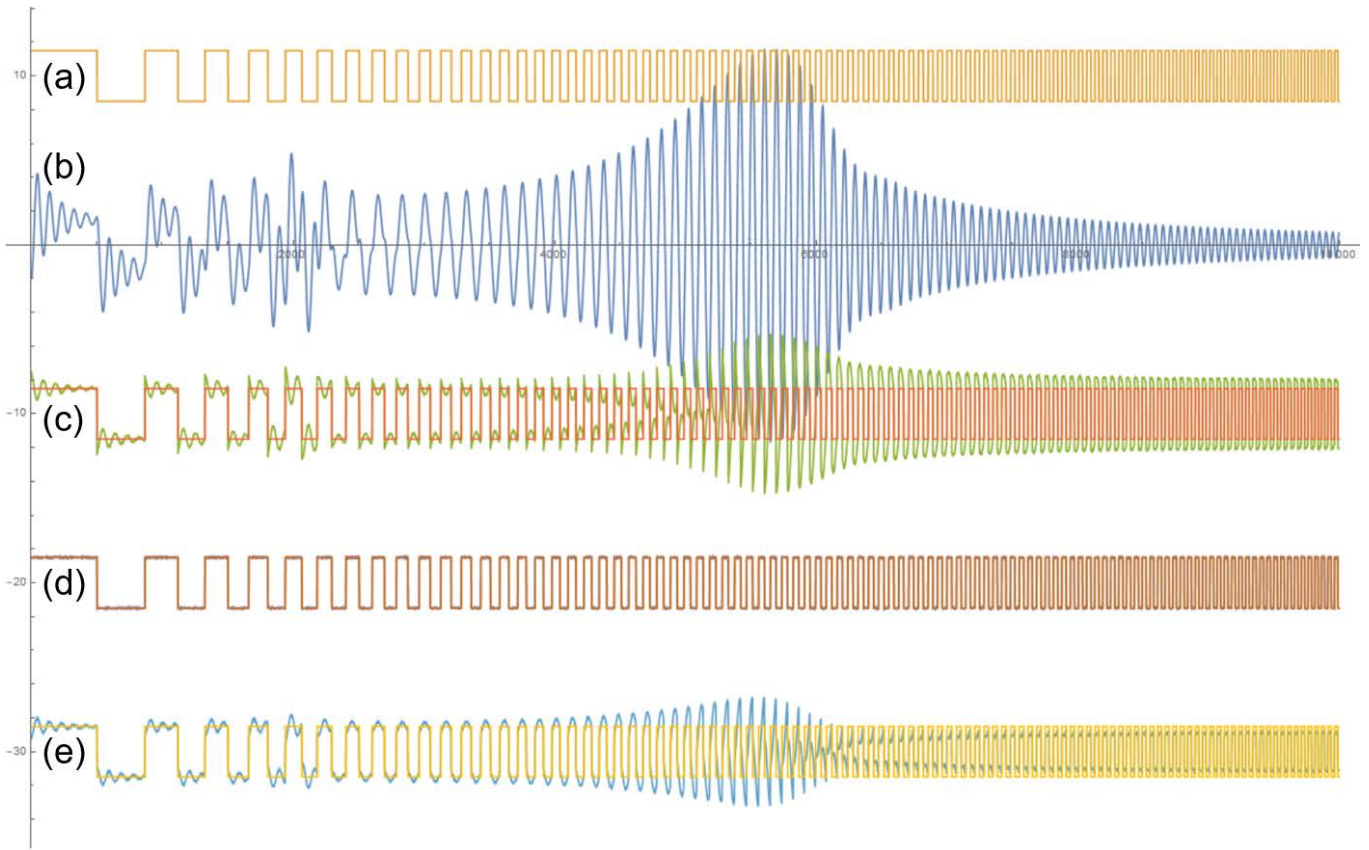

Fig. 7. Numerical simulation of the real-time waveform recovery in the scheme of Fig. 3(c) for a system with three-pole transfer characteristics shown in Fig. 1(b), tested with a frequency-varying rectangular input waveform. (a) Input waveform $x_i(t)$. (b) Response of the system output $x_o(t)$ calculated by numerically solving the differential equation with the two initial conditions $x_o(0)$ and $\dot{x}_o(0)$ arbitrarily chosen. (c) Compensated waveform $x_r(t)$ (overlaid with the input waveform $x_i(t)$) calculated from the waveform in (b) with all the $\gamma_i$ parameters intentionally offset from their optimal values by replacing every $T$ by $0.9T$. (d) Compensated waveform $x_r(t)$ (overlaid with the input waveform $x_i(t)$) calculated from the waveform in (b) with all the $\gamma_i$ parameters set at their optimal values. (e) Compensated waveform $x_r(t)$ (overlaid with the input waveform $x_i(t)$) calculated from the waveform in (b) with all the $\gamma_i$ parameters intentionally offset from their optimal values by replacing every $T$ by $1.1T$. Note that the strong transient waveforms after the sharp transitions and the strong resonant waveform near $4000 < t < 7000$ are exactly cancelled out in (d).

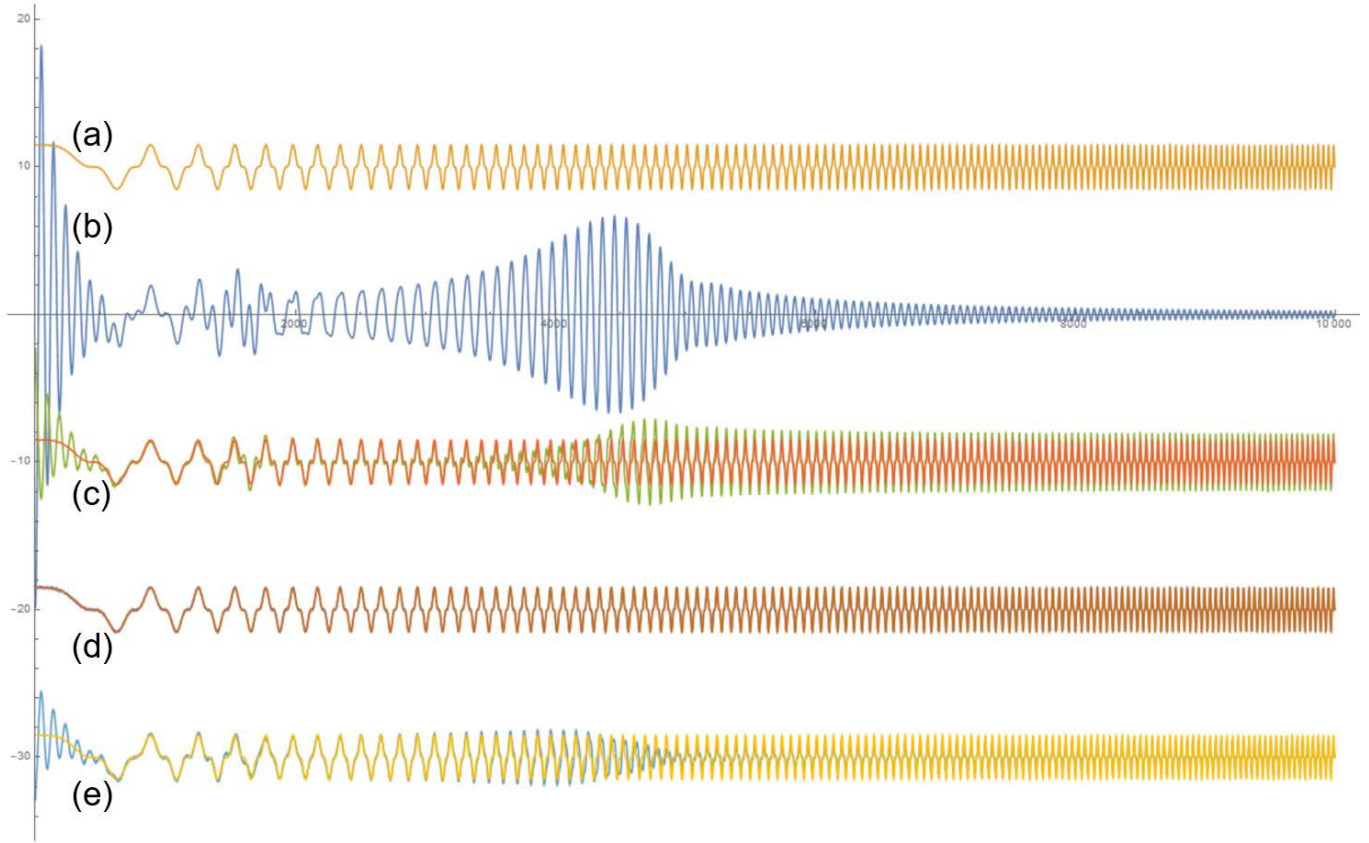

Fig. 8. Numerical simulation of the real-time waveform recovery in the scheme of Fig. 3(c) for a system with three-pole transfer characteristics shown in Fig. 1(b), tested with a frequency-varying cosine-cubed input waveform. (a) Input waveform $x_i(t)$. (b) Response of the system output $x_o(t)$ calculated by numerically solving the differential equation with the two initial conditions $x_o(0)$ and $\dot{x}_o(0)$ arbitrarily chosen. (c) Compensated waveform $x_r(t)$ (overlaid with the input waveform $x_i(t)$) calculated from the waveform in (b) with all the $\gamma_i$ parameters intentionally offset from their optimal values by replacing every $T$ by $0.9T$. (d) Compensated waveform $x_r(t)$ (overlaid with the input waveform $x_i(t)$) calculated from the waveform in (b) with all the $\gamma_i$ parameters set at their optimal values. (e) Compensated waveform $x_r(t)$ (overlaid with the input waveform $x_i(t)$) calculated from the waveform in (b) with all the $\gamma_i$ parameters intentionally offset from their optimal values by replacing every $T$ by $1.1T$. Note that the strong transient waveform near $0 < t < 800$ in (b) and the strong resonant waveform near $3500 < t < 5500$ are exactly cancelled out in (d).

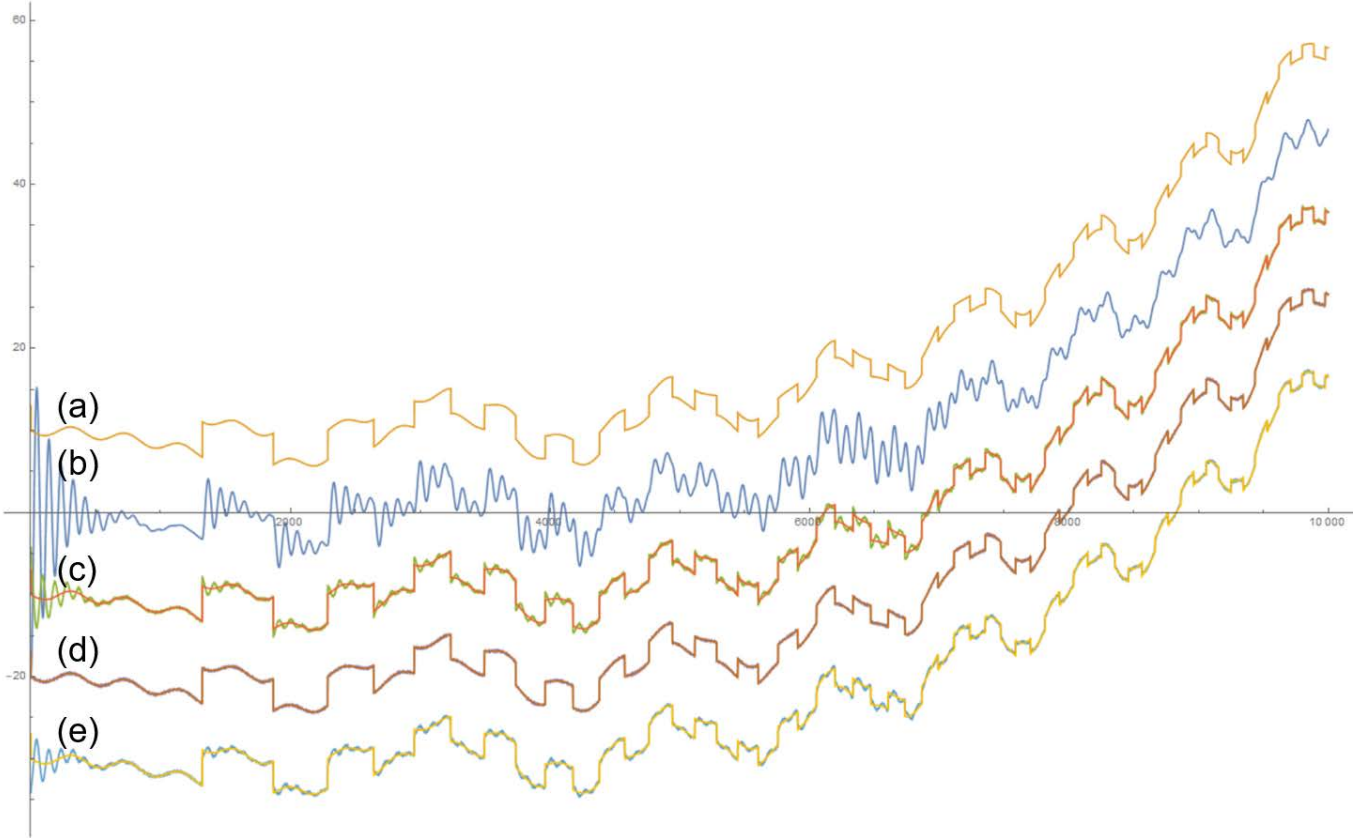

Fig. 9. Numerical simulation of the real-time waveform recovery in the scheme of Fig. 3(c) for a system with three-pole transfer characteristics shown in Fig. 1(b), tested with an aperiodic input waveform. (a) Input waveform $x_i(t)$. (b) Response of the system output $x_o(t)$ calculated by numerically solving the differential equation with the two initial conditions $x_o(0)$ and $\dot{x}_o(0)$ arbitrarily chosen. (c) Compensated waveform $x_r(t)$ (overlaid with the input waveform $x_i(t)$) calculated from the waveform in (b) with all the $\gamma_i$ parameters intentionally offset from their optimal values by replacing every $T$ by $0.9T$. (d) Compensated waveform $x_r(t)$ (overlaid with the input waveform $x_i(t)$) calculated from the waveform in (b) with all the $\gamma_i$ parameters set at their optimal values. (e) Compensated waveform $x_r(t)$ (overlaid with the input waveform $x_i(t)$) calculated from the waveform in (b) with all the $\gamma_i$ parameters intentionally offset from their optimal values by replacing every $T$ by $1.1T$. Note that the strong transient waveforms after the sharp transitions are exactly cancelled out in (d) and the recovery is DC-accurate even with the parameters slightly offset from their optimal values as shown in (c)-(e).